\documentclass[aps,prd,showpacs,superscriptaddress,nofootinbib,twocolumn]{revtex4} 
\usepackage{graphicx}
\usepackage{amsmath}
\usepackage{natbib}
\begin{document}
\title{High-energy neutrinos from reverse shocks in choked and successful relativistic jets} 

\author{Shunsaku Horiuchi}
\affiliation{Department of Physics, School of Science, The University
  of Tokyo, Tokyo 113-0033, Japan}

\author{Shin'ichiro Ando}
\affiliation{California Institute of Technology, Mail Code 130-33,
  Pasadena, CA 91125}

\date{March 18th, 2008}

\begin{abstract}
Highly relativistic jets are a key element of current gamma-ray burst models, where the jet kinetic energy is converted to radiation energy at optically thin shocks. High-energy neutrinos are also expected, from interactions of protons accelerated in the same shocks. Here we revisit the early evolution of a relativistic jet, while the jet is still inside the star, and investigate its neutrino emission. In particular we study propagation of mildly relativistic and ultrarelativistic jets through a type Ib progenitor, and follow reverse shocks as the jets cross the star. We show that protons can be accelerated to $10^4$--$10^5$ GeV at reverse shocks, and efficiently produce mesons. The mesons experience significant cooling, suppressing subsequent neutrino emission. We show, however, that the neutrino yield from the reverse shock is still reasonably large, especially for low-luminosity and long-duration jets, where meson cooling is less severe. We discuss implications of our results in the context of neutrinos from choked jets, which are completely shock heated and do not break out of the star. From a choked jet with isotropic equivalent energy of $10^{53}\,\mathrm{erg}$ at 10 Mpc, we expect $\sim$20 neutrino events at IceCube.
\end{abstract}

\maketitle
\section{Introduction} 

The leading model of gamma-ray bursts (GRBs) involves a relativistic ($\Gamma_j\gtrsim 100$, where $\Gamma_j$ is the jet Lorentz factor) fireball jet, where the observed gamma rays are produced by radiation from Fermi-accelerated electrons in optically thin shocks (for reviews see e.g.~\cite{Zhang:2003uk,Piran:2004ba,Meszaros:2006rc}). In the most popular model for the more common long-duration GRBs, the so-called collapsar model, the core of a massive star collapses to a black hole or neutron star, driving a highly relativistic jet which breaks out of the star~\cite{Woosley:1993wj,MacFadyen:1998vz}. Within this scenario, high-energy neutrinos from photomeson interactions of accelerated protons have been studied~\cite{Waxman:1997ti,Rachen:1998fd,Waxman:1998yy,Waxman:1999ai,Bahcall:1999yr,AlvarezMuniz:2000st,Meszaros:2001ms,Guetta:2002du,Dermer:2003zv,Li:2002dw,Murase:2006mm}. With $1\,\mathrm{km^3}$ size neutrino detectors such as IceCube~\cite{Ahrens:2003ix} and KM3Net~\cite{Katz:2006wv} being constructed, we are entering an age when we can test various predictions for high-energy neutrinos. Waxman and Bahcall predicted $\epsilon_\nu > 10^5\,\mathrm{TeV}$ neutrinos from external reverse shocks~\cite{Waxman:1999ai}, and similar predictions have been made in Refs.~\cite{Dermer:2003zv,Li:2002dw} for the forward shock. In addition, internal shocks can occur while the relativistic jet is still in the star. M\'esz\'aros and Waxman predicted a $\epsilon_\nu \gtrsim 5 \,\mathrm{TeV}$ neutrino precursor burst from such a set up~\cite{Meszaros:2001ms}. It is expected that a precursor neutrino burst is present even if the jet does not successfully break out of the star, producing a choked jet that is dark in gamma rays and bright in neutrinos.

Inferring from the observed rate of GRBs, highly relativistic ($\Gamma_j \gtrsim 100$) jets occur in perhaps $\lesssim 10^{-3}$ of core-collapse supernovae~\cite{Berger:2003kg}. However, the jet signature may be more common, and a significantly higher portion of supernovae may be endowed with slower, mildly relativistic jets; according to late-time radio observations, perhaps as high as a few $\%$~\cite{Totani:2003rk,Berger:2003kg,Soderberg:2004ma,Granot:2004rh,van Putten:2004dh}. The recent detection of low-luminosity GRB 060218 also suggests from detection rates that low-luminosity GRBs, with mildly relativistic ejecta, are more common than conventional high-luminosity GRBs by factors as large as $10^2$~\cite{Campana:2006qe,Cobb:2006cu,Pian:2006pr,Soderberg:2006vh,Liang:2006ab,Toma:2006iu,Waxman:2007rr}. The occurrence rate is high enough that despite their lower luminosity, their contribution to the neutrino background is comparable to or larger than conventional GRBs~\cite{Murase:2006mm,Gupta:2006jm}.

In addition to being more frequent, mildly relativistic jets are more baryon rich, which work positively for neutrino emission. Indeed, the detectability of neutrinos from internal shocks of mildly relativistic jets has been studied by Razzaque, M\'esz\'aros, and Waxman~\cite{Razzaque:2004yv} and extended by Ando and Beacom~\cite{Ando:2005xi} (see also~\cite{Razzaque:2005bh,Koers:2007je}). In these scenarios, internal shocks occur at $r_s\approx 10^{10}$--$10^{11}\,\mathrm{cm}$, smaller than the typical radius of type Ib progenitors. Accelerated protons interact via proton-photon ($p\gamma$) and proton-proton ($pp$) interactions to produce mesons, which consequently produce neutrinos on decay.

In this paper we investigate neutrinos arising from protons that are accelerated in the reverse shock, while the relativistic jet is still propagating inside the star. We adopt a type Ib progenitor as our model star, but we discuss how our results apply to type Ic and II progenitors also. The reverse shock is strongly motivated because it is a generic prediction of the initial stages of any jet moving through the star, regardless of its Lorentz factor, and whether it breaks out of the star or not. Therefore, we consider two types of supernovae: one is the perhaps more numerous supernova containing a mildly relativistic ($\Gamma_j \sim 10$) jet, and the other is a supernova containing an ultrarelativistic ($\Gamma_j \sim 100$) jet that can ultimately cause the observed GRBs. In both cases we consider choked as well as successful jets, depending on its duration.

We show that protons can be accelerated to $10^4$--$10^5$ GeV in the reverse shock. The low and high-energy protons produce mesons efficiently by $pp$ and $p\gamma$ interactions respectively. For energies in between, cooling mechanisms are faster, and result in a characteristic suppression feature in the meson spectrum. Mesons that are produced experience significant cooling due to the high proton and photon densities in the shocked jet. However, we still expect a reasonable number of neutrino events. For instance, for a choked mildly relativistic jet with isotropic energy $10^{54}$ erg at 10 Mpc, we expect $\sim2$ neutrino events at IceCube from the reverse shock. Interestingly, assuming the same jet Lorentz factor, we find that a lower isotropic energy of $10^{53}$ erg produces more neutrino events ($\sim20$ events). This is due to meson cooling. Specifically, we find that neutrinos are favored by high-Lorentz factor, low-luminosity, and long-duration jets, as meson cooling is less significant for such jets. The rate of core-collapse supernova within 10 Mpc is expected to be $\sim$1--3 yr$^{-1}$~\cite{Ando:2005ka}, implying particle acceleration in reverse shocks can potentially be tested in the next decades. 

Neutrino emission from the reverse shock is potentially important for short-duration jets\footnote{We note that the word ``short-duration'' here does not mean another class of GRBs, which is often referred to as short-duration and hard-spectrum GRB, and is believed to be associated with compact-star mergers~\cite{Ando:2004pc,Nakar:2007yr}.} that are completely shocked and do not break out of the star, i.e., choked jets. We argue that while it is possible that internal shocks occur prior to jet shocking (in particular for low $\Gamma_j$ jets),  the allowed range of radii for internal shocks to occur is constrained from above by the reverse shock, and below by the fact that the stellar material is opaque to high-energy neutrinos. This suggests that neutrino emission from choked jets are of similar magnitude to our estimates.

This paper is organized as follows. In Sec.~II we discuss the evolution of the jet as it propagates through the star. In Sec.~III we treat proton acceleration, and suppression of neutrino emission due to proton and meson cooling. In Sec.~IV we utilize the results of the previous sections to investigate the detectability of high-energy neutrinos from reverse shocks. Finally, we finish with discussions in Sec.~V and conclusions in Sec.~VI. 

\section{Jet Models}

\subsection{Jet dynamics}

\begin{figure}[t]
\includegraphics[width=3.4in,clip=true]{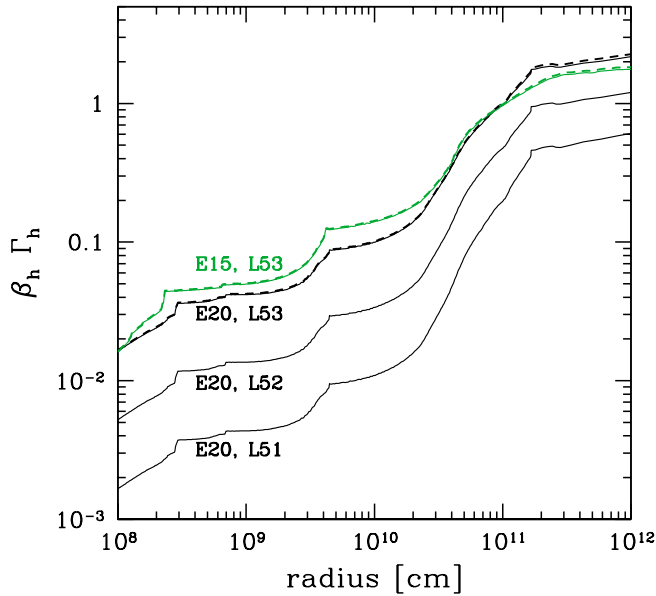}
\caption{\label{velocity}Jet head velocity $\beta_h \Gamma_h$ as a
  function of radius. E15 and E20 are two progenitor models of mass 
  15$M_\odot$ and 20$M_\odot$ respectively, as detailed in the text. 
  L51--L53 denotes the isotropic equivalent luminosity of the jet, 
  $L_{\rm{iso}}=10^{51}$--$10^{53}\,\mathrm{erg \, s^{-1}}$. For the 
  L53 case, two jet Lorentz factors are plotted: mildly relativistic
  ($\Gamma_j=10$, solid) and ultrarelativistic ($\Gamma_j=100$, dashed). 
  For L52 and L51, mildly relativistic jets are shown.}
\end{figure}

We first discuss the evolution of a relativistic $\Gamma_j \gg 1$ jet as it strikes the stellar matter. The aim is to obtain the evolution of the jet head and reverse shock velocities. There are several numerical simulations on the propagation of an initially hot jet in the literature~\cite{Aloy:1999ai,Zhang:2002yk,Zhang:2003rp,Umeda:2005ce,Mizuta:2006tz}, as well as analytic treatments of initially cold jets~\cite{Meszaros:2001ms,Meszaros:2001vr,Waxman:2002uu,Matzner:2002ti}. We follow and extend the latter. Generally, two shocks form when a jet strikes external matter: a reverse shock that decelerates the head of the jet to a Lorentz factor $\Gamma_h$, and a forward shock that accelerates the external material to $\Gamma_h$. The jet and stellar material moving at $\Gamma_h$ are separated by a contact discontinuity, and are in pressure balance. Equations governing the evolution of the shocks are (for $\Gamma_j \gg 1$)~\cite{Blandford:1976uq,Sari:1995nm} 
\begin{eqnarray} \label{shockequations}
e_s/n_s m_p c^2 = \Gamma_h-1, \quad n_s/n_{ext}=4\Gamma_h+3, \nonumber
\\
e_h/n_h m_p c^2 = \bar{\Gamma}_h-1, \quad n_h/n_j=4\bar{\Gamma}_h+3, 
\end{eqnarray}
where $m_p$ is the proton mass and $c$ is the speed of light. Thermodynamic quantities $n_i$ and $e_i$ (particle number density and internal energy) are measured in the fluids' rest frames, and we have used the following notations: $j$ (jet, unshocked), $h$ (jet head, shocked), $s$ (stellar, shocked), and $ext$ (stellar, unshocked). Lorentz factors are measured in the lab frame, except for $\bar{\Gamma}_h$ which is the Lorentz factor of the jet head measured in the frame of the unshocked jet, 
\begin{equation} \label{gamma3bar}
\bar{\Gamma}_h = \Gamma_j \Gamma_h (1-\beta_j \beta_h),
\end{equation}
where $\beta_i$ is the velocity as a fraction of $c$. The jet particle density is given by (for a jet with constant opening angle)
\begin{equation} \label{n_j}
n_j(r)=\frac{L_{\rm{iso}}}{4 \pi r^2 \Gamma_j^2 m_p c^3},
\end{equation}
where $L_{\rm{iso}}$ is the isotropic jet luminosity. We obtain $\Gamma_h$ by equating the pressure of the shocked jet head $p_h = e_h/3$ and the pressure of the shocked stellar matter $p_s = e_s/3$~\cite{Meszaros:2001ms}. From Eq.~(\ref{shockequations}), this requires us to solve
\begin{equation}\label{v_h}
\frac{n_j}{n_{ext}} =
\frac{(4\Gamma_h+3)(\Gamma_h-1)}
     {(4\bar{\Gamma}_h+3)(\bar{\Gamma}_h-1)},
\end{equation}
by substitution of Eqs.~(\ref{gamma3bar}) and (\ref{n_j}). In what follows we use $n_{ext}$ from the progenitor model E20 from Ref.~\cite{Heger:1999ax}, an initially $20M_\odot$ rotating star that is reduced to $11M_\odot$ by mass-loss. It has a $1.7M_\odot$ Fe core of radius $r_{\rm{Fe}}\sim2\times10^8\,\mathrm{cm}$, a $7.7M_\odot$ He core that extends to $r_{\rm{He}}\sim2\times10^{11}\,\mathrm{cm}$, and a surrounding H envelope. However, the results that we present here apply qualitatively to other models of different masses, with or without H envelopes. As an illustration of this point we include other progenitor models in Fig.~\ref{velocity}.
 
Due to rotation, the mass along the rotational axis becomes substantially lower than the equatorial, and the rotational axis is an easy escape route for a jet. We inject jets at a radius $r \sim 2 \times 10^8\,\mathrm{cm}$ along the rotational axis, which corresponds to an enclosed mass of $1.6 M_\odot$. Since the explosion and jet mechanisms are still under debate, they are treated parametrically with the following parameters: the isotropic equivalent energy $E_{\rm{iso}}$ and duration $T_j$ (which yield the isotropic equivalent luminosity through $L_{\rm{iso}} = E_{\rm iso} / T_j$), Lorentz factor $\Gamma_j$, and variability $t_v$. We consider two types of jets, (i) mildly relativistic jets with $\Gamma_j=10$, and (ii) ultrarelativistic jets with $\Gamma_j=100$.

In Fig.~\ref{velocity} we show the jet head velocity $\beta_h$ multiplied by $\Gamma_h$ as a function of radius, obtained from solving Eq.~(\ref{v_h}). For $L_{\rm{iso}}=10^{53}\,\mathrm{erg \, s^{-1}}$, we show our two jet types $\Gamma_j=10$ and $\Gamma_j=100$, for two progenitor models. The first progenitor model is E20 as described above, and the second is E15, a similarly rotating star but with initial mass $15 M_\odot$~\cite{Heger:1999ax}. We see that the evolution of $\beta_h$ is independent of $\Gamma_j$, and that the initial jet head velocity is always subrelativistic. The latter is because the external density is very much larger than the jet density, $n_{ext} \gg n_j$, for conceivable set of jet parameters. As the jet continues to propagate, the stellar density falls faster with radius ($n_{ext} \propto r^{-3}$) than the jet density ($n_j\propto r^{-2}$), and the jet head effectively accelerates, reaching mildly relativistic values at $r \gtrsim r_{\rm{He}} \sim$ a few $\times 10^{11}\,\mathrm{cm}$. Although the initial value of $\beta_h$ scales somewhat with $L_{\rm{iso}}$, the radius at which the velocity reaches $c$ is only weakly dependent on jet parameters. This is because the jet head accelerates to relativistic velocities when the external stellar density suddenly drops, at the edge of the He core. In fact, rewriting Eq.~(\ref{v_h}) in the relativistic limit ($\Gamma_j,\Gamma_h \gg 1$) yields $\Gamma_h \propto L_{\rm{iso}}^{1/4}r^{-1/2}n_{ext}^{-1/4}$~\cite{Meszaros:2001ms}.

Note that the kinks in Fig.~\ref{velocity} corresponding to composition changes in the progenitor model are model dependent, e.g., an abundance of oxygen resides at $r\sim 4\times 10^{9}\,\mathrm{cm}$. However, the general features of $\beta_h$ that concern us, including its initially $\beta_h \ll 1$ value and its reaching $\sim 1$ at $r \sim r_{\rm{He}}$, are model independent. We therefore consider only the E20 progenitor model in the rest of our paper.

\begin{table} [t]
\caption{\label{table:times}Summary of choked jet characteristic
  times. Shown are the core crossing time $t_{\rm{He}}$, and the jet
  duration time $T_\times$ required for reverse shock crossing to
  occur at $r_\times=5\times 10^{10}$ cm. These values apply to both mildly
  relativistic and ultrarelativistic jets.}
\begin{ruledtabular}
\begin{tabular}{ccccc}
$L_{\rm{iso}}$ [erg/s] & $t_{\rm{He}}$ [s] & $T_\times$ [s] &
  $E_{\rm{iso}}$ [erg] & Model \\ \hline 
$10^{53}$ & 17 & 12 & $1\times 10^{54}$ & A \\
$10^{52}$ & 43 & 34 & $3\times 10^{53}$ & B \\
$10^{51}$ & 120 & 100 & $1\times 10^{53}$ & C \\
\end{tabular}
\end{ruledtabular}
\end{table} 

From the velocity profile we now obtain an estimate of the time taken for the jet to propagate through the stellar core, $t_{\rm{He}}\sim 10$--$100$ seconds for $L_{\rm{iso}}=10^{51}$--$10^{53}\,\mathrm{erg \,s^{-1}}$ (see Table \ref{table:times}). We adopt model B in Table~\ref{table:times} as our fiducial supernova jet model.

\subsection{Choked and successful jets}

Since jets reach relativistic velocities as they leave the He core, the time taken for a jet to cross the H envelope $t_{\rm{H}} \sim r_{\rm{H}}/(2c\Gamma_h^2) \ll t_{\rm{He}}$~\cite{Meszaros:2001vr}, and the time taken to leave the star is approximately the time taken to cross the core, $t_{\rm{He}} \approx 10$--$100$ seconds. If the jet duration $T_j$ is less than $t_{\rm{He}}$, the cold jet will be unable to break out of the star, i.e., a choked jet that is dark in gamma rays. On the other hand, if $T_j > t_{\rm{He}}$, the cold jet can break out, and upon entering the H envelope or interstellar medium, receive a large velocity boost. A GRB is then possible.

We first discuss choked jets that do not break out of the star. Such jets are progressively shocked and decelerated by the reverse shock, until at a crossing radius we define as $r_\times$, the entire jet is shocked. The reverse shock is relativistic in the jet frame ($\bar{\Gamma}_h \gg 1$), and efficiently converts the jet bulk kinetic energy to internal energy. Energy conversion is thus complete in just one jet crossing by the reverse shock~\cite{Sari:1995nm,Piran:2004ba}.

Internal shocks between jet shells of different luminosity and $\Gamma_j$ will occur at radii no larger than $r_s \approx 2\Gamma_j^2 c \, t_v$. If they occur before the jets are decelerated significantly by the reverse shock, a fraction of the jet's bulk kinetic energy is converted at internal shocks~\cite{Kobayashi:1997jk,Daigne:2003tp}. Therefore, the relative sizes of $r_s$ and $r_\times$ indicate at which shock (internal or reverse) the jet's kinetic energy is converted. In the current work we focus on cases where $r_\times < r_s$, i.e., the jet's kinetic energy is converted by the reverse shock. We discuss in Sec.~V internal shocks occurring before reverse shock crossing.

\begin{table}
\caption{\label{table:jets}Definitions for jets and their
  observables. Successful highly relativistic jets are the leading
  model for GRBs. The successful mildly relativistic jet and the
  choked jets are dark in gamma rays, because shocks typically occur
  while the jet is optically thick. Entry ``this work'' represents the
  specific focus of this work. In addition, since the reverse shock
  occurs in successful jets too, the present work applies to the
  entire neutrino ($\nu$) row.}
\begin{ruledtabular}
\begin{tabular}{ccccc}
 &\multicolumn{2}{c}{Choked}&\multicolumn{2}{c}{Successful}\\
 & $\Gamma_j=10$ & $\Gamma_j=100$ & $\Gamma_j=10$ & $\Gamma_j=100$ \\
 \hline
$\gamma$ rays & dark & dark & typically dark & bright \\
$\nu$ & this work & this work & \cite{Razzaque:2004yv,Ando:2005xi} &
 \cite{Meszaros:2001ms,Razzaque:2003uv} \\
\end{tabular}
\end{ruledtabular}
\end{table} 

To estimate $r_\times$ for a choked jet, we note that $\bar{\Gamma}_h \approx \Gamma_j$ during propagation in the core. The reverse shock thus crosses the entire length of the jet in a time $\Delta/c \sim T_j$, where $\Delta$ is the jet length in the lab frame. We obtain $r_\times$ as the distance the jet head has traveled in time $T_j$. As an example, adopting a jet luminosity $L_{\rm{iso}} = 10^{52}\,\mathrm{erg\,s^{-1}}$, we obtain $t_{\rm{He}}\sim 43$ s; a jet with duration $T_j=30\,\mathrm{s}$ is therefore choked, and we find that $r_\times \approx 4 \times 10^{10} \,\mathrm{cm}$ for both the mildly relativistic as well as ultrarelativistic jet cases. For calculation purposes we set $r_\times=5\times 10^{10}\,\mathrm{cm}$. This requires a jet duration of $T_\times \sim 0.7$--$0.8 t_{\rm{He}}$, depending on the value of $L_{\rm{iso}}$. We summarize duration times in Table \ref{table:times}.

We note that we could have in principle chosen a smaller $r_\times$ due to smaller $T_\times$. However, as we discuss in Sec.~V, the stellar matter becomes opaque to high-energy neutrinos for radii $r \lesssim 10^{10}$ cm, and small radii are uninteresting. We also note here that rotational effects decrease the matter density along the rotational axis, increasing the jet head velocity at small radii and thereby shortening the core crossing time $t_{\rm{He}}$. A decrease in density also decreases neutrino opacity. However, the effect is not expected to be large, because density modifications typically occur at radii smaller than or equal to $r_{\rm{Fe}}$~\cite{Meszaros:2001vr}.

Next we discuss jets that successfully break out of the star, with $T_j > t_{\rm{He}}$. These jets are progressively shocked as they move through the progenitor, but the reverse shock does not cross the entire jet. Therefore, for successful jets, only the front portion of the jet is shocked. We assume $r_\times$ just inside of the He core radius. We stress that for successful jets, $r_\times$ is \emph{not} the point of reverse shock crossing, but the last point of shocking by the reverse shock.

A few words about the jet models must be noted. First, we have assumed a conical jet geometry in Eq.~(\ref{n_j}). This is a good approximation for jets propagating in a sufficiently sparse medium, such as the outer regions of stars. From analytic studies, the jet opening angle $\theta \propto r^{-1/2}$ for $r \lesssim r_{\rm{Fe}}$ and $\theta \propto$ constant for $r \gtrsim r_{\rm{Fe}}$~\cite{Meszaros:2001vr}. We have therefore injected jets at a radius $\sim r_{\rm{Fe}}$ with profile Eq.~(\ref{n_j}). On the other hand, under conditions such as propagation in the dense iron core and/or when strong magnetic fields are associated with the launch of the jet, the high external matter density and/or magnetic hoop stress work to collimate the jet. Numerical studies find that jets remain tightly collimated in the iron core~\cite{Burrows:2007yx}, and further beyond the stronger the magnetic field~\cite{Takiwaki:2004kf}. For jets that remain strongly collimated, the reverse shock becomes increasingly less important due to the more penetrating nature of the jet. Further, it is less likely for these jets to form choked jets. For illustration, consider an extreme case where a jet of luminosity $10^{52}$ erg $\rm{s^{-1}}$ remains tightly cylindrical all the way until break out, with constant cylindrical radius $10^7$ cm~\cite{Takiwaki:2004kf}. For such a jet, we find that $t_{\rm{He}} \sim$ 6 s and $T_\times \lesssim 2$ s.

Second, we have assumed a type Ib progenitor. A jet with $L_{\rm iso} = 10^{52}$ erg s$^{-1}$ propagating through a type Ic progenitor would clear the core in $t_{\rm{CNO}}\sim 20$ seconds, and jets are more likely to break out.

Third, we have defined successful jets as those which break out of the star. However, these do not necessarily always produce observable GRBs---for small $\Gamma_j$ and $t_v$, internal shocks will occur while the jet is optically thick to photons. For example, $r_s \sim 10^{12} \Gamma_{j,1}^2t_{v,-1}$ cm (we define $Q_\alpha = Q / 10^\alpha$ for a quantity $Q$ in cgs units), while the jet typically becomes optically thin to photons at $[(L_{\rm{iso}} \sigma_T)/(4 \pi \Gamma_j m_p c^2)]^{1/2} \sim 10^{14}L_{\rm{iso},52}^{1/2}\Gamma_{j,1}^{-1/2}$ cm, where $\sigma_T$ is the Thomson scattering cross section. The jet is still transparent to high-energy neutrinos. We summarize our various jet definitions in Table~\ref{table:jets}.

To conclude, we use the following crossing radii: (i) $r_\times = 5 \times 10^{10} \,\mathrm{cm}$ for choked jets and (ii) $r_\times = 10^{11} \,\mathrm{cm}$ for successful jets.

\section{Neutrino Production}

In this section we discuss proton acceleration and neutrino production through decays of pion and kaon mesons. The reverse shock is likely to be collisionless, as we discuss in section IIIA. If first-order Fermi acceleration is realized, a fraction of the jet protons are accelerated, with a spectrum
\begin{equation} \label{protonspectrum}
\frac{dn_p}{d\epsilon_p}=\frac{U_p}
  {\int^{\epsilon_{p,\rm{max}}}_{\epsilon_{p,\rm{min}}}
  d\epsilon_p \epsilon_p  \frac{dn_p}{d\epsilon_p}} \epsilon_p^{-p},
\end{equation}
where the power-law index $p \approx 2.3$ in the ultrarelativistic shock limit~\cite{Achterberg:2001rx,Keshet:2004ch}, and $p = 2$ for a nonrelativistic shock. In this paper, we optimistically adopt $p = 2$ for all of our figures, but we present numerical results for $p=$ 2.1, 2.3, and 2.5 also. The minimum proton energy would be $\epsilon_{p,\rm{min}} \sim \Gamma_j m_p c^2$, but otherwise it is not well known; we assume $\epsilon_{p,\rm{min}} = 10\,\mathrm{GeV}$, but its exact value does not affect our results.

\subsection{Maximum proton energy}

\begin{figure}[t]
\includegraphics[width=3.4in,clip=true]{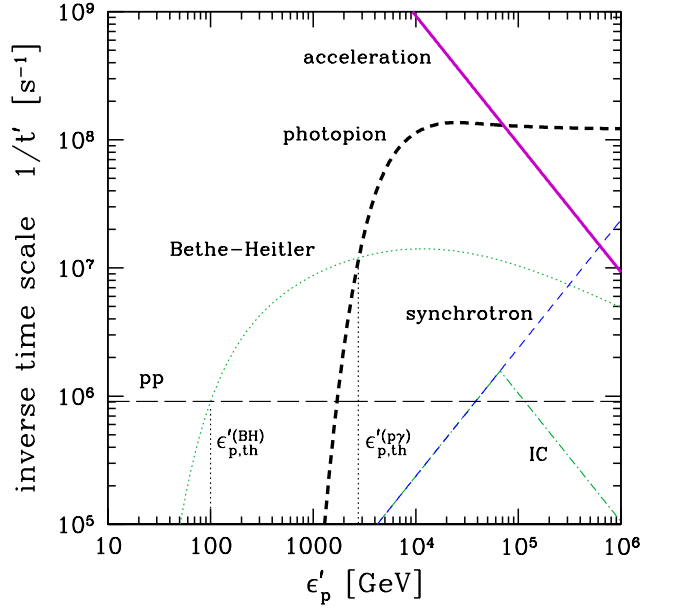}
\caption{\label{cooling}Inverse of proton cooling and acceleration time scales, in the shocked jet head frame, as functions of the proton energy. The case of reverse shock crossing at $r_\times=5\times10^{10}\,\mathrm{cm}$ is plotted for supernova jet model B, with other parameters $\Gamma_j=10$, $\varepsilon_e=\varepsilon_B=0.1$, and $t_v=10^{-1}\,\mathrm{s}$. Cooling mechanisms shown are: synchrotron (blue, thin dashed), inverse-Compton (green, dot-dashed), Bethe-Heitler (green, dotted), proton-proton (black, long dashed), and photopion (black, thick dashed). Also labeled are threshold energies $\epsilon_{p,th}^{\prime (BH)}$ and $\epsilon_{p,th}^{\prime (p\gamma)}$ (see Sec.~IIIB). Photopion limits the proton energy to $7 \times 10^4\,\mathrm{GeV}$.}
\end{figure}

The maximum proton acceleration energy is determined by comparing the proton's acceleration time scale to its energy loss time scales. Ando and Beacom have shown that the photopion process is the most competitive cooling mechanism in the proton energy range of interest~\cite{Ando:2005xi}. We therefore focus on photopion production, and refer the reader to Appendix~A for treatments of other cooling mechanisms such as synchrotron cooling. We will however for completeness display all cooling mechanisms in Fig.~\ref{cooling}. Note that the Bethe-Heitler process turns out to be an important cooling process for $pp$ interactions, as we will discuss in the next subsection. 

The energy density in the shocked jet head is $e_h^\prime = (4\bar{\Gamma}_h+3)(\bar{\Gamma}_h-1)n_jm_pc^2 \approx (\bar{\Gamma}_h/\Gamma_j)^2 L_{\rm{iso}}/(\pi r^2 c)$. We have specifically used primes ($\prime$) to remind that these quantities are in the jet head comoving frame. We assume that a fraction $\varepsilon_e = \varepsilon_B = 0.1$ of the shocked plasma internal energy is converted to relativistic electrons and magnetic fields, in analogy to GRBs. This gives the comoving magnetic field strength as
\begin{eqnarray} \label{Bfield}
B^\prime&\approx&[(8 \varepsilon_B L_{\rm{iso}}) / (r^2 c)]^{1/2}
(\bar{\Gamma}_h/\Gamma_j) \nonumber \\
&\approx& 1 \times 10^{10} \, \left(
\frac{\varepsilon_{B,-1}^{1/2}L_{\rm{iso},52}^{1/2}}
{r_{\times,10.7}} \right)
\left( \frac{\bar{\Gamma}_{h,1}}{\Gamma_{j,1}} \right) \, \mathrm{G},
\end{eqnarray}
where we have assumed $\bar{\Gamma}_h \approx \Gamma_j$ since $\Gamma_h \approx 1$ for the jet scenarios studied here. One can obtain from Eq.~(\ref{gamma3bar}) that for example, $\Gamma_h=2$ yields $\bar{\Gamma}_h/\Gamma_j \approx 0.3$. The relativistic electrons immediately lose energy by synchrotron radiation, and the photons thermalize due to the large opacity $n^\prime_e \sigma_T \Delta_h \sim 10^{6}$, where $n^\prime_e = (4\Gamma_j+3)n_j(r_\times) \approx 3 \times 10^{21}\,\mathrm{cm^{-3}}$ and $\Delta_h$ is the proper thickness of the shocked jet plasma. Here we adopt $\Delta_h = 0.2 \theta r$, where $\theta = 10^{-1}$~\cite{Meszaros:2001ms}. The black-body temperature is then
\begin{eqnarray} \label{photons}
T^\prime_r &\approx& [(15 \varepsilon_e L_{\rm{iso}}
  \hbar^3 c^3)/(\pi^3 r^2 c)]^{1/4} 
(\bar{\Gamma}_h/\Gamma_j)^{1/2} \nonumber \\
&\approx& 13 \, \left(
\frac{\varepsilon_{e,-1}^{1/4}L_{\rm{iso},52}^{1/4}}
{r_{\times,10.7}^{1/2}} \right) 
\left( \frac{\bar{\Gamma}_{h,1}}{\Gamma_{j,1}} \right)^{1/2} \,\mathrm{keV},
\end{eqnarray}
and the average photon energy is $\bar{\epsilon}_\gamma^\prime \approx 2.7 T_r^\prime$. The average density is
\begin{eqnarray} \label{photons2}
\bar{n}^\prime_\gamma &\approx& 0.33 [(\varepsilon_e L_{\rm{iso}}) / (\pi^3 r^2 c^2 \hbar) ]^{3/4}
(\bar{\Gamma}_h/\Gamma_j)^{3/2} \nonumber \\
&\approx& 7 \times 10^{25} \, \left(
\frac{\varepsilon_{e,-1}^{3/4}L_{\rm{iso},52}^{3/4}}
{r_{\times,10.7}^{3/2}} \right) 
\left(\frac{\bar{\Gamma}_{h,1}}{\Gamma_{j,1}} \right)^{3/2} \,\mathrm{cm^{-3}}.
\end{eqnarray}

The total proton cooling time is determined by the inverse sum $t^{\prime-1}_{c,\rm{tot}}=\Sigma_i \, t^{\prime-1}_{c,i}$, where $i$ runs through the various cooling mechanisms. As discussed in the opening paragraph, we focus on cooling by photopion production. Protons interact with the dense photon field Eq.~(\ref{photons2}), and for sufficient proton-photon centre of inertia energies mesons are produced. We determine the cooling time scale by $t_{p\gamma}^\prime=\epsilon_p^\prime/(c\sigma_{p\gamma} n_{\gamma}^\prime \Delta\epsilon_p^\prime)$. We assume the conventional inelasticity $K=\Delta\epsilon_p^\prime/\epsilon_p^\prime=[1-(m_p^2-m_\pi^2)/s]/2$ where $s$ is the invariant mass of the system. We fit $\sigma_{p\gamma}$ to~\cite{Eidelman:2004wy}, and use a blackbody photon spectrum of temperature Eq.~(\ref{photons}). The time scale is plotted in Fig.~\ref{cooling}.

The acceleration time scale in the jet head frame is given by $t^\prime_{\rm{acc}} \equiv \theta_F R^\prime_L / c$, where $\theta_F$ is a constant if we assume the diffusion coefficient is proportional to the Bohm diffusion coefficient, and $R^\prime_L$ is the Larmor radius. $\theta_F \ge 10$ is a fairly conservative value, and $\theta_F \sim 1$ can be achieved for mildly relativistic shocks~\cite{Rachen:1998fd}. We use $\theta_F \sim 10$ which gives agreement with previous studies~\cite{Razzaque:2004yv,Ando:2005xi}. Using the magnetic field in Eq.~(\ref{Bfield}), we obtain
\begin{eqnarray}
t^\prime_{\rm{acc}} &\approx& \theta_F\epsilon^\prime_p/(e B^\prime c)
\nonumber \\
&\approx& 1 \times 10^{-13} \, \left(
\frac{r_{\times,10.7} \, \epsilon^\prime_{p,0}}
{\varepsilon_{B,-1}^{1/2}L_{\rm{iso},52}^{1/2}} \right) 
\left( \frac{\bar{\Gamma}_{h,1}}{\Gamma_{j,1}} \right)^{-1}
\, \mathrm{s},
\end{eqnarray}
where $\epsilon^\prime_p$ is measured in GeV. 

In Fig.~\ref{cooling}, we plot the proton acceleration and cooling time scales as functions of the proton energy, both in the jet head comoving frame. Acceleration is stronger than cooling at low proton energies, but photopion cooling becomes stronger at higher energies. It is evident that photopion cooling is the strongest cooling mechanism. Equating $t^\prime_{\rm{acc}} = t^\prime_{p,p\gamma}$, we obtain the maximum proton energy as $\epsilon_{p,\rm{max}}^\prime \approx 7 \times 10^4 \,\mathrm{GeV}$. The parameter dependency is $r^{1/2}_{\times,10.7}\varepsilon_{B,-1}^{1/2}\varepsilon_{e,-1}^{-3/4}L_{\rm{iso},52}^{-1/4}(\bar{\Gamma}_{h,1}/\Gamma_{j,1})^{-1/2}$ in the flat region of $t^\prime_{p\gamma}$. 

We note that the reverse shock is likely to be collisionless, because the proton plasma frequency $\omega_p\sim 10^{14}\,\mathrm{Hz}$ is larger than the typical radiation and particle collision frequencies. For example, the frequency of proton-photon collision is $\omega_{\rm{coll}}\sim n^\prime_\gamma \sigma_{p\gamma}c\sim 10^{10}\,\mathrm{Hz}$, while the frequency of proton-proton collision is $\sim n^\prime_p \sigma_{pp}c\sim 10^{6}\,\mathrm{Hz}$.

\subsection{Proton cooling and meson spectrum}
Depending on its energy, protons produce mesons via $pp$ and/or $p\gamma$ interactions. We consider the pion and kaon mesons. The multiplicity in each $pp$ and $p\gamma$ interaction is taken to be $1$ for pions and $0.1$ for kaons; this matches the required ratio in the relevant energies~\cite{Kass:1979nf,Alner:1985zc,Lindsey:1991pt}. Generally, kaons produce higher energy neutrinos~\cite{Ando:2005xi,Asano:2006zzb}, but we additionally find that they can dominate those from pions because of less efficient kaon cooling.

For low energy protons, $pp$ interactions dominate. Meson production by $pp$ interaction is efficient since the opacity is very high, $n^\prime_p \sigma_{pp} \Delta_h \sim 10^5$, where $n^\prime_p = (4\Gamma_j+3)n_j(r_\times) \approx 3 \times 10^{21}\,\mathrm{cm^{-3}}$ and $\sigma_{pp}=5 \times 10^{-26}\,\mathrm{cm^2}$~\cite{Eidelman:2004wy}. Proton cooling by pair production (Bethe-Heitler, hereafter BH, see Appendix for treatment) overtakes $pp$ interactions at higher energies (see Fig.~\ref{cooling}). We define this transition as $\epsilon^{\prime(BH)}_{p,th}$. At even higher energies $p\gamma$ dominates over BH cooling, and we define this transition energy as $\epsilon^{\prime(p\gamma)}_{p,th}$. The $p\gamma$ interaction is efficient at producing mesons since the opacity is $\bar{n}_\gamma^\prime \sigma_{p\gamma}\Delta_h \sim 10^7$, where we use $\sigma_{p\gamma}=10^{-28}\,\mathrm{cm^2}$~\cite{Eidelman:2004wy}. For model B in Table I, $\epsilon^{\prime(BH)}_{p,th} \approx 100$ GeV and $\epsilon^{\prime (p\gamma)}_{p,th} \approx 2800$ GeV. 

We assume mesons are produced with 20\% of the parent proton energy. In the absence of competing cooling mechanisms, the meson spectrum follows the initial proton spectrum. On the other hand, when other cooling processes are faster than meson-yielding $pp$ and $p\gamma$ interactions, the meson spectrum is suppressed by an amount equal to the ratio of time scales. We find that the Bethe-Heitler process is a competitive cooling mechanism for $pp$ interactions, and the resulting meson spectrum will be suppressed by a factor given as 
\begin{equation} 
\zeta_{BH} = \left\{ 
\begin{array}{ll}
\frac{t_{BH}^\prime}{t_{pp}^\prime} & \quad \mathrm{for}\,\,
\epsilon_{p,th}^{\prime(BH)} < \epsilon_p^\prime < \epsilon_{p,th}^{\prime(p\gamma)} \\
1 & \quad \mathrm{otherwise}.
\end{array}\right.
\end{equation}
We here approximate $t^{\prime}_{BH}$ as a power-law $t^{\prime}_{BH} = t^{\prime}_{pp}(\epsilon_p^\prime / \epsilon_{p,th}^{\prime(BH)})^{-k}$ for $\epsilon_{p,th}^{\prime(BH)} < \epsilon_p^\prime < \epsilon_{p,th}^{\prime (p\gamma)}$, where $k$ is a constant. We find that the value of $k$ is fairly model independent, with values $k \approx 0.8$ (1.2) for $\Gamma_j$ = 10 (100) for model B.

\subsection{Meson cooling and neutrino spectrum}
In this section we discuss meson cooling and the neutrino spectrum. Mesons cool similarly to protons, by radiative and hadronic (collisions with protons) processes. The meson cooling time scales are analogous to those for the proton and can be summarized as $t^\prime_{\rm{rad}} = (3m^4c^3) / [4\sigma_Tm_e^2\epsilon^\prime(U^\prime_\gamma+U^\prime_B)]$ and $t^\prime_{\rm{had}}=\epsilon^\prime/(c\sigma_h n^\prime_p \Delta \epsilon^\prime)$, where $m$ and $\epsilon^\prime$ are the meson mass and energy, and $U^\prime_\gamma=\epsilon^\prime_\gamma n^\prime_\gamma$ and $U^\prime_B=B^{\prime2}/(8\pi)$ are the photon and magnetic field energy densities (we assume inverse-Compton process is in the Thompson regime; synchrotron dominates at high energies). Here, $\sigma_h=5 \times 10^{-26}\,\mathrm{cm^2}$~\cite{Eidelman:2004wy} is the cross section for meson-proton collisions and $\Delta \epsilon^\prime=0.8 \epsilon^\prime$~\cite{Brenner:1981kf} is the energy lost by the meson per collision. The total cooling time is $t^{\prime -1}_c=t^{\prime -1}_{\rm{rad}} + t^{\prime -1}_{\rm{had}}$, and the time scales are
\begin{equation} \label{mesonhadronic}
t^\prime_{\pi,\rm{had}}\approx 3 \times 10^{-7}\, \left(
\frac{ r_{\times,10.7}^{2} \Gamma_{j,1}}{L_{\rm{iso},52}}
\right) \, \mathrm{s}
\end{equation}
and
\begin{eqnarray}
t^\prime_{\pi,\rm{rad}} &\approx& 1 \times 10^{-5} \left(
\frac{r_{\times,10.7}^2}
{(\varepsilon_{e,-1} + \varepsilon_{B,-1}) L_{\rm{iso},52} \,\epsilon^\prime_{\pi,0}} \right)  \nonumber \\
&&
\times \left( \frac{\bar{\Gamma}_{h,1}}{\Gamma_{j,1}} \right)^{-2}
\, \mathrm{s}
\end{eqnarray}
for pions. For kaons, the radiative cooling time scale is longer, because $t^\prime_{\rm{rad}}\propto m^4$. Hadronic cooling dominates at low proton energies, and radiative cooling becomes important at higher energies.

Charged pions and kaons decay into neutrinos through $\pi^\pm,K^\pm \to \mu^\pm + \nu_\mu (\bar{\nu}_\mu)$, with the muon neutrino taking 1/4 of the meson energy. We neglect secondary neutrinos from muon decays since muons immediately lose energy by radiative cooling. We define the break energy $\epsilon^{\prime(1)}_{\rm{brk}}$ by equating $\gamma^\prime \tau = t^\prime_c \sim t^\prime_{\rm{had}}$, where $\gamma^\prime$ and $\tau$ are the meson Lorentz factor and proper lifetime. Below $\epsilon^{\prime(1)}_{\rm{brk}}$, neutrinos have a typical energy $0.05\epsilon_p^\prime$ and a flat power per decade neutrino spectrum. Above $\epsilon^{\prime(1)}_{\rm{brk}}$, the spectrum is suppressed by a factor $t^\prime_{\rm{had}}/(\gamma^\prime \tau)$. Next, we define $\epsilon^{\prime(2)}_{\rm{brk}}$ when radiative cooling begins to dominate, by equating $t^\prime_{\rm{had}} = t^\prime_{\rm{rad}}$. Due to relativistic effects, the neutrino energy in the observer frame is related to the parent meson energy in the jet head frame as $\epsilon_\nu = \Gamma_h \epsilon^\prime_\pi/4$ and $\epsilon_\nu = \Gamma_h \epsilon^\prime_K/2$. We define the suppression function $\zeta$ as
\begin{eqnarray} \label{suppression}
\zeta(\epsilon_\nu) = 
\left\{
\begin{array}{lll}
1 & \quad \mathrm{for}\,\,\epsilon_\nu < \epsilon^{(1)}_{\nu,\rm{brk}}
\\
\epsilon^{(1)}_{\rm{brk}} / \epsilon_\nu & \quad
\mathrm{for}\,\,\epsilon^{(1)}_{\nu,\rm{brk}} \le \epsilon_\nu <
\epsilon^{(2)}_{\nu,\rm{brk}} \\
\epsilon^{(1)}_{\rm{brk}} \epsilon^{(2)}_{\rm{brk}} / \epsilon_\nu^{2}
& \quad \mathrm{for}\,\,\epsilon_\nu \ge
\epsilon^{(2)}_{\nu,\rm{brk}}.
\end{array} \right.
\end{eqnarray}
For some supernova jet parameters, it is possible that meson goes from decay dominated straight to radiation cooling dominated. For these cases we define the break energy $\epsilon^{\prime(r)}_{\rm{brk}}$ by equating $\gamma^\prime \tau = t^\prime_{\rm rad}$, and the suppression
function $\zeta$ as
\begin{eqnarray}
\zeta(\epsilon_\nu) = 
\left\{
\begin{array}{ll}
1 & \quad \mathrm{for}\,\,\epsilon_\nu < \epsilon^{(r)}_{\nu,\rm{brk}}
\\
( \epsilon^{(r)}_{\rm{brk}} / \epsilon_\nu )^{2}
&\quad\mathrm{for}\,\, \epsilon_\nu \ge
\epsilon^{(r)}_{\nu,\rm{brk}}.
\end{array} \right.
\end{eqnarray}

The break energies for neutrinos from pion decay, in a choked mildly relativistic jet, are
\begin{equation} \label{break12}
\epsilon^{\pi(1)}_{\nu,\rm{brk}}=0.4 \,\mathrm{GeV} \quad ; \quad
\epsilon^{\pi(2)}_{\nu,\rm{brk}}= 10 \,\mathrm{GeV}.
\end{equation}
The parameter dependencies are $r^2_{\times,10.7}\Gamma_{j,1} \Gamma_{h,0} L_{\rm{iso},52}^{-1}$ and $\Gamma_{h,0} (\varepsilon_{e,-1}+\varepsilon_{B,-1})^{-1} \Gamma_{j,1}^{-1} (\bar{\Gamma}_{h,1}/\Gamma_{j,1})^{-2}$ for the first and second break energies, respectively. On the other hand, break energies for neutrinos from kaon decay are higher due to the higher kaon mass and faster decay time;
\begin{equation} \label{break34}
\epsilon^{K(1)}_{\nu,\rm{brk}}= 5 \,\mathrm{GeV} \quad ; \quad
\epsilon^{K(2)}_{\nu,\rm{brk}}= 3000 \,\mathrm{GeV}.
\end{equation}
The parameter dependencies are the same as for pions. The break energies are smaller than those of internal shocks~\cite{Ando:2005xi} with the same jet parameters, by as much as  $\sim$2 orders of magnitude. This is a combination of the higher proton and photon densities in the shocked jet causing more effective cooling, and the fact that the neutrino energies in the shocked jet head receive no Lorentz boosting.

Next, the break energies for a choked ultrarelativistic jet are
\begin{eqnarray}
&&\epsilon^{\pi(1)}_{\nu,\rm{brk}}= 4 \,\mathrm{GeV} \quad ; \quad
\epsilon^{\pi(2)}_{\nu,\rm{brk}}= 0.9 \,\mathrm{GeV}, \\
&&\epsilon^{K(1)}_{\nu,\rm{brk}}= 60 \,\mathrm{GeV} \quad ; \quad
\epsilon^{K(2)}_{\nu,\rm{brk}}= 300 \,\mathrm{GeV},
\end{eqnarray}
which can also be obtained from scaling Eqs.~(\ref{break12})--(\ref{break34}) by their dependencies on $\Gamma_j$
Importantly for neutrino emission, suppression due to radiative cooling is independent of $\Gamma_j$, since the product of $\epsilon^{(1)}_{\nu,\rm{brk}} \times \epsilon^{(2)}_{\nu,\rm{brk}}$ is unchanged. Also, note that $\epsilon^{\pi(1)}_{\nu,\rm{brk}} > \epsilon^{\pi(2)}_{\nu,\rm{brk}}$, indicating that radiative cooling dominates the cooling process for pions. In this case,
\begin{equation}
\epsilon^{\pi(r)}_{\nu,\rm{brk}} = 2 \, \left( 
\frac{r_{\times,10.7}\Gamma_{h,0}}
{(\varepsilon_{e,-1}+\varepsilon_{B,-1})^{1/2}L_{\rm{iso},52}^{1/2}} \right)
\left( \frac{\bar{\Gamma}_{h,2}}{\Gamma_{j,2}} \right)^{-2} \,\mathrm{GeV},
\end{equation}

\begin{figure}[t]
\includegraphics[width=3.4in,clip=true]{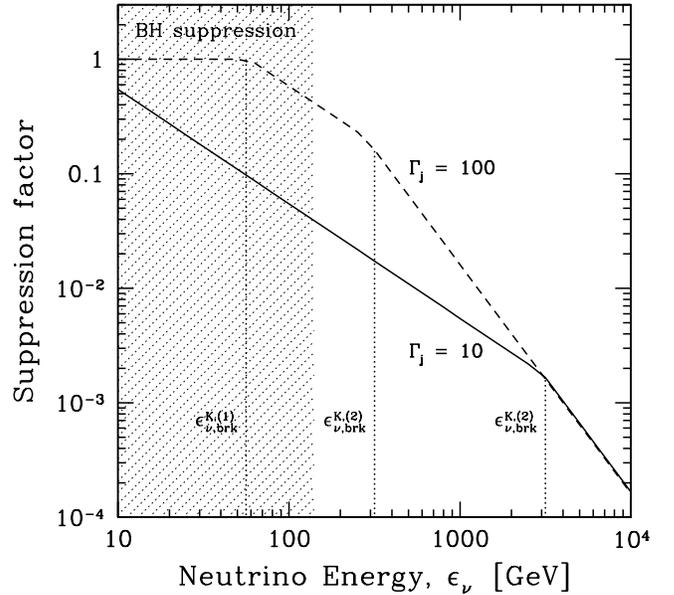}
\caption{\label{suppression1}Suppression factor as a function of neutrino energy, for neutrinos from kaon decays. Shown are the choked mildly relativistic jet (solid lines) and choked ultrarelativistic jet (dashed lines). The kaon experiences less cooling for an ultrarelativistic jet. Note that pions are always in the radiation cooling regime for neutrino energies shown. We also shade in for reference the energies where Bethe-Heitler cooling is competitive and suppresses the meson spectrum.}
\end{figure}

We show in Fig.~\ref{suppression1} the resulting suppression factors for neutrinos from kaons, for our mildly relativistic and ultrarelativistic jets. Suppression is less intense for relativistic jets. We also shade in for reference the energy range where the meson spectrum is suppressed by Bethe-Heitler cooling, from 5--140 GeV. Note that for pions, radiative cooling dominates for energies of interest.

\section{Detection of Neutrino Bursts}

\begin{figure}[t]
\includegraphics[width=3.4in,clip=true]{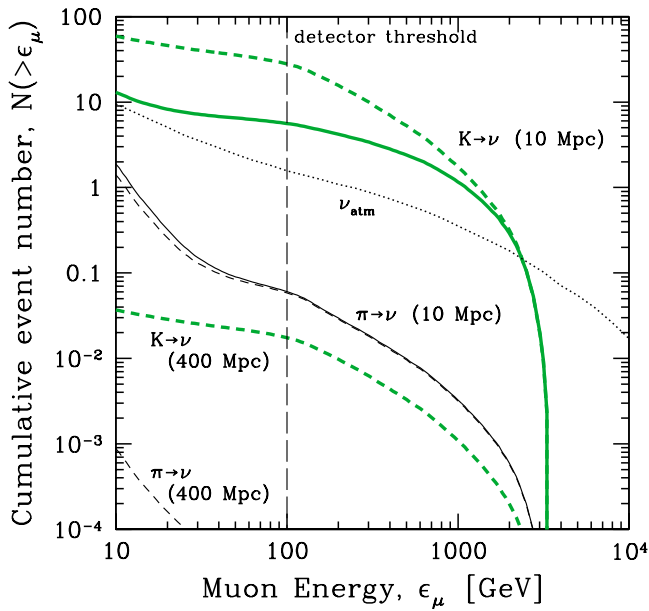}
\caption{\label{muon}Cumulative muon event number from choked jets with model B parameters. Thin-black (thick-green) lines denote pion (kaon) contributions. The majority originates from kaons. Solid lines denote a mildly relativistic jet and a supernova distance of $10\,\mathrm{Mpc}$. Dashed lines denote an ultrarelativistic jet and two supernova distances - $10\,\mathrm{Mpc}$ for illustrative purposes and $400\,\mathrm{Mpc}$ for event calculations, as labeled. Due to weaker kaon cooling, an ultrarelativistic jet yields more neutrinos than a mildly relativistic one. The spectrum becomes almost flat around $\sim$ 100 GeV, due to Bethe-Heitler suppression of the meson spectrum. The total number of events expected above the detector threshold energy is about 6 for the mildly relativistic case. For the ultrarelativistic case with distance $400\,\mathrm{Mpc}$ we expect 0.02 events. Also shown is the atmospheric neutrino background over 1 day in a $3^\circ$ circle.}
\end{figure}

In this section we estimate the emission and detectability of high-energy neutrinos from reverse shock accelerated protons. We consider a supernovae at a distance of $D_L$, and use the code ANIS (All Neutrino Interaction Generator)~\cite{Gazizov:2004va} to calculate the neutrino induced muon spectrum at a neutrino detector with an effective area of $1\,\mathrm{km^2}$, i.e.~an IceCube class detector utilizing upgoing muons. First, we estimate the fluence of neutrinos as~\cite{Ando:2005xi}
\begin{equation}\label{fluence}
F_\nu = \frac{\langle n \rangle B_\nu}{\kappa} \frac{E_{\rm iso}}{4\pi D_L^2 \ln (\epsilon_{p,{\rm max}}^\prime /
   \epsilon_{p,{\rm min}}^\prime )}
\frac{\zeta (\epsilon_\nu)\zeta_{BH} (\epsilon_\nu)}{\epsilon_\nu^2}, 
\end{equation} 
where $\langle n \rangle$ is the meson multiplicity ($1$ for pions and $0.1$ for kaons), $B_\nu$ is the branching ratio of meson decay into neutrinos ($1$ for pions and $0.6$ for kaons), and $\kappa^{-1}$ is the fraction of proton energy carried by neutrinos in the absence of energy loses; $1/8$ for pions and $1/4$ for kaons, since neutral and charged mesons are produced with roughly equal probability, and muon neutrinos carry roughly $1/4$ ($1/2$) of the pion (kaon) energy in meson decay. The functions $\zeta_{BH}(\epsilon_\nu)$ and $\zeta(\epsilon_\nu)$ are the suppression factors due to proton and meson coolings, respectively, and the $\mathrm{ln} (\epsilon^{\prime}_{p,\rm{max}} / \epsilon^{\prime}_{p,\rm{min}})$ factor normalizes the proton spectrum to the jet energy (for $p=2$).

In Fig.~\ref{muon} we show the cumulative spectrum of neutrino induced muons, from a supernova with a choked jet. We show three cases - a supernova distance of $D_L=10\,\mathrm{Mpc}$ possessing a mildly relativistic $\Gamma_j=10$ jet (solid lines), a supernova distance of $D_L=10\,\mathrm{Mpc}$ possessing an ultrarelativistic $\Gamma_j=100$ jet (dashed lines), and a similarly ultrarelativistic jet but supernova distance $D_L=400\,\mathrm{Mpc}$ (dashed lines labeled 400 Mpc). We use model B with parameters $\varepsilon_e=\varepsilon_B=0.1$ and $t_v=10^{-2}\,\mathrm{s}$. We take into account the muon range, which effectively increases the detector volume. We evaluate the muon energy $\epsilon_\mu$ when it enters the detector if it is produced outside, or at the production point if it is inside. Considering a detection threshold muon energy of $100\,\mathrm{GeV}$, we expect about 6 muon events from a 10 Mpc supernova possessing a choked mildly relativistic jet. Similarly, we expect about 28 from an ultrarelativistic jet. We have plotted the ultrarelativistic jet case at an optimistic distance of 10 Mpc to illustrate the effect of meson cooling. Although the pion contributions are similar between the two jets, the kaon contribution is larger in the ultrarelativistic case due to the weaker suppression. In the remainder we assume the distance 400 Mpc for ultrarelativistic jet calculations. For model B, this yields 0.02 events. 

Note that since the muon spectrum is steep, a lower detector energy threshold will yield much higher events. Additionally, one may observe the suppression due to Bethe-Heitler cooling, which is expected to manifest below $0.05 \epsilon_{p,th}^{\prime (p\gamma)} \sim 140$ GeV. For a steeper initial proton spectrum, motivated from studies of ultrarelativistic shocks, the expected event is 4 (2, 1) for $p=2.1$ (2.3, 2.5). In all jets, the majority of neutrinos come from kaons, because kaons are more massive than pions and suffer less cooling (the radiative cooling time scale is $\propto m^4$). Also, the kaon decay time is slightly shorter than the pion. The events discussed here arrive in a $\approx$ 34 second time bin and $\sim$ 3$^\circ$ angular bin, allowing very strong rejection of atmospheric neutrino backgrounds. We show in the figure the atmospheric neutrino background within $1$ day and a $3^\circ$ bin for comparison. The expected neutrino signal from a choked mildly relativistic jet easily exceeds the atmospheric neutrino background.

\begin{figure}[t]
\includegraphics[width=3.4in,clip=true]{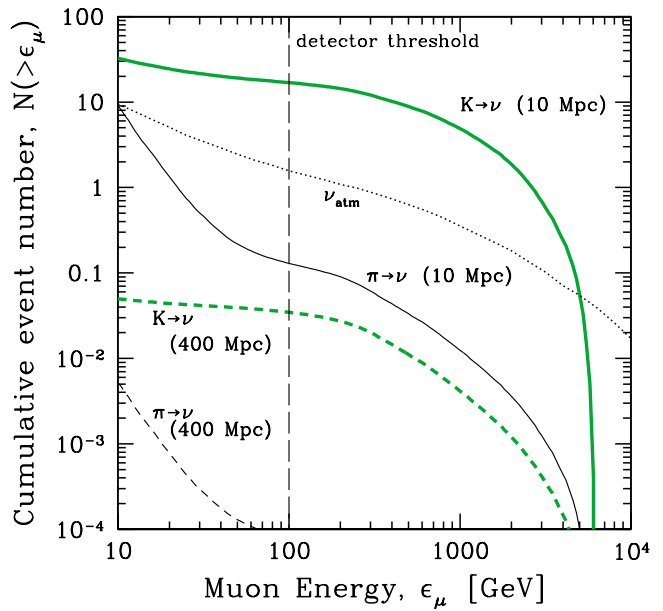}
\caption{\label{muonC}Same as Fig.~\ref{muon} but for model C parameters. Approximately 17 (0.04) events are expected above the detector threshold for a single  mildly relativistic (ultrarelativistic) jet and supernova distance $10\,\mathrm{Mpc}$ ($400\,\mathrm{Mpc}$).}
\end{figure}

\begin{figure}[t]
\includegraphics[width=3.4in,clip=true]{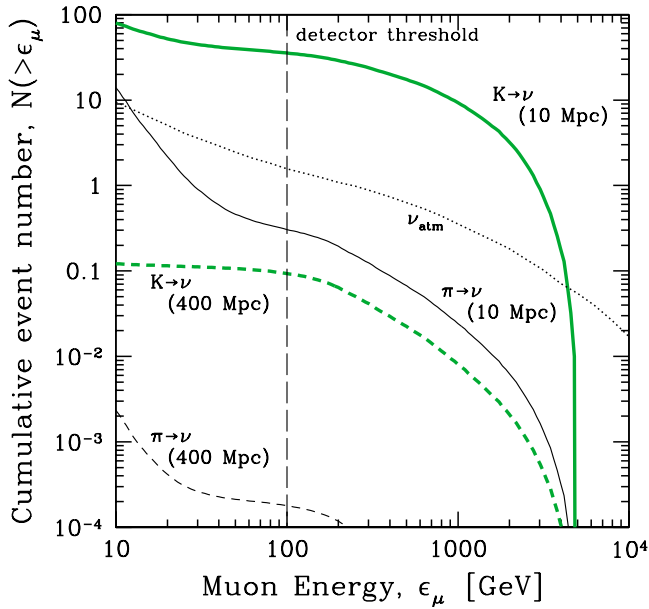}
\caption{\label{muon2}Same as Fig.~\ref{muon} but for jets with successful jets. The supernova and jet parameters are $L_{\rm{iso}}=10^{52}\, \mathrm{erg\,s^{-1}}$, $T_j=100\,\mathrm{s}$, and $t_v=0.1\,\mathrm{s}$. About 36 (0.1) events are expected above the detector threshold muon energy from a mildly relativistic (ultrarelativistic) jet supernova at a distance $10\,\mathrm{Mpc}$ ($400\,\mathrm{Mpc}$) away.}
\end{figure}
 
Next we discuss a scan of supernova and jet parameters. First, as shown above, a larger $\Gamma_j$ reduces kaon suppression and increases neutrino emission. Next, we consider jet model A of Table~\ref{table:times}, i.e., with larger $L_{\rm{iso}}$ but smaller $T_j$, and an overall larger $E_{\rm{iso}}$. Since the neutrino fluence scales with $E_{\rm{iso}}$, one naively expects more neutrinos. However, the expected muon event rate is in fact smaller, $2$ ($0.01$) for a mildly relativistic (ultrarelativistic) jet. This is because of stronger meson cooling: the proton and photon densities in the shocked jet scale as $n^\prime_p \propto L_{\rm{iso}}$ and $n^\prime_\gamma \propto L_{\rm{iso}}^{3/4}$, so a larger $L_{\rm{iso}}$ results in stronger meson cooling. We conclude that neutrino emission from choked jets are more favorable from low-luminosity and long-duration jets, even if $E_{\rm{iso}}$ is smaller (assuming the same jet Lorentz factor). Indeed, model C predicts 17 (0.04) events at IceCube. We show the cumulative muon event spectrum for jet model C in Fig.~\ref{muonC}.

In Fig.~\ref{muon2} we show the same as Fig.~\ref{muon} but for successful jets. We show a mildly relativistic $\Gamma_j=10$ jet at $10\,\mathrm{Mpc}$ (solid lines), and an ultrarelativistic $\Gamma_j=100$ jet at $400\,\mathrm{Mpc}$ (dashed lines). The other parameters used are $L_{\rm{iso}}=10^{52}\,\mathrm{erg\,s^{-1}}$, $T_j=100\,\mathrm{s}$, $t_v=0.1\,\mathrm{s}$, $\varepsilon_e=\varepsilon_B=0.1$, and $r_\times=10^{11}\,\mathrm{cm}$. With these parameters, $t_{\rm{He}}\sim 43$ s, and the cold jet does indeed break out of the star. We introduce an additional factor of 1/2 to the neutrino fluence for successful jets, because only about $\sim t_{\rm{He}}/T_j \sim 1/2$ of the jet length is shocked by the reverse shock. Note that our chosen parameters have the equivalent $E_{\rm{iso}}$ to jet model A. However, we expect $\sim(1/2)(2^2/10^{-1})\sim20$ times more neutrino events than jet model A, because $L_{\rm{iso}}$ is a factor 10 smaller and $r_\times$ is a factor 2 larger. Indeed, evaluating the muon event rate yields about 36 (0.1) events per mildly relativistic (ultrarelativistic) jet. As in the case of choked jets, most of the contribution comes from kaons. These events will show temporal and angular clustering, and in addition, these will appear as precursor signals to GRB photons, allowing very strong rejection of background. 

\section{Discussion}

Our result is that neutrinos from the reverse shock are strongly suppressed due to proton and meson cooling. However, we still expect a reasonable number of events at neutrino detectors. One reason for this is that proton cooling affects only energies near and below current detector threshold energies. This produces a suppression feature in the spectrum but does not decrease event counts at IceCube. For our models A, B, and C in Table~\ref{table:times}, we expect 2, 6, and 17 muon events respectively at a IceCube class detector. Here, we have calculated for a supernova $10\,\mathrm{Mpc}$ away, possessing a $\Gamma_j=10$ mildly relativistic jet, and a $p=2$ initial proton spectrum. If we assume a steeper index, $p=$ 2.1 (2.3, 2.5), we expect 4 (2, 1) events for jet model B. In all cases, the neutrino flux is dominated by decay of kaons, since cooling is less efficient for kaons than pions. 

Since neutrino emission from the reverse shock is suppressed by meson cooling, we find that neutrino emission is favored by parameter sets for which meson cooling is weaker -- high-Lorentz factor, low-luminosity, and long-duration jets. This fact is readily observable from the neutrino event predictions of models A, B, and C. As further illustration, if we assume a $\Gamma_j=100$ ultrarelativistic jet we expect 11, 28, and 55 events from jet models A, B, and C respectively. Note that supernova and jet parameters are expected to show considerable scatter. Hence, given neutrino emission is cooling suppressed, the event count need not necessarily correlate with the total burst energy. 

Neutrino emission from the reverse shock is particularly important for choked jets, which are completely shock decelerated by the reverse shock. Is it possible for choked jets to produce more neutrino emission than our above estimates? If the duration is long enough such that $r_s < r_\times$, it is possible that internal shocks accelerate protons in the unshocked part of the jet, while the jet head is being shocked. The internal shock accelerated protons do not travel far, producing mesons efficiently via $pp$ interactions with the large target jet proton density~\cite{Razzaque:2004yv,Ando:2005xi}. For high-Lorentz factor jets, the protons can travel and interact with termination shock photons~\cite{Meszaros:2001ms}. In both cases meson cooling is less severe than in the reverse shock, and we can expect more neutrino emission. However, internal shocks must occur at large enough radii that the stellar material is not opaque to high-energy neutrinos. The neutrino opacity is $\tau_\nu = N_{ext}(r) \sigma_{\nu p}(\epsilon_\nu)$, where $N_{ext}$ is the column number density of the star and $\sigma_{\nu p}$ is the sum of neutrino charged-current and neutral-current cross sections, which is approximately $\propto \epsilon_\nu$ in the energy range in question. For a $10^2\,\rm{GeV}$ neutrino emitted at $r=10^{10}\,\rm{cm}$, $N_{ext} \approx 8\times 10^{35}\,\mathrm{cm^{-2}}$ and $\sigma_{\nu p} = 9 \times 10^{-37}\,\mathrm{cm^2}$~\cite{Gandhi:1998ri}, and we find that $\tau_\nu \approx 0.68$. For a $10^3\,\rm{GeV}$ neutrino, $\tau_\nu \approx 6$. Thus, only neutrinos with energies $\epsilon_\nu < 10^2 \,\mathrm{GeV}$ can leave the star from $r=10^{10}\,\mathrm{cm}$. Clearly, these results depend on the stellar density profile assumed, and rotational effects have been neglected in our simple estimate. However, it seems clear that the stellar matter is opaque to high-energy neutrinos for small radii, constraining the range of allowed radii for internal shocks.

\begin{figure}[t]
\includegraphics[width=3.4in,clip=true]{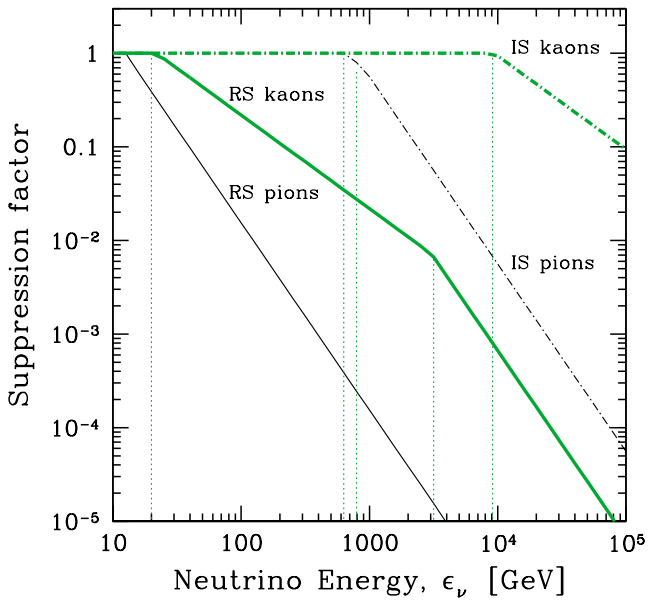}
\caption{\label{suppression2}Suppression factor for neutrinos from the reverse shock (solid lines) and the internal shock (dot-dashed lines). We adopt our successful mildly relativisitc jet ($L_{\rm{iso}}=10^{52}\,\mathrm{erg\,s^{-1}}$, $T_j=100\,\mathrm{s}$, $t_v=0.02\,\mathrm{s}$), and show the suppression factors for pions (thin-black lines) and kaons (thick-green lines). The vertical dotted lines indicate break energies.}
\end{figure}

The case for the reverse shock in successful jets, which break out of the star, seem better. This however is not due to any intrinsic property of successful jets. It is merely because for a given total jet energy, successful jets have longer durations and lower luminosities compared to choked jets. This results in less meson cooling; for example, we expect 2 events from a choked jet (of energy $E_{\rm iso} = 10^{54}\,\mathrm{erg}$), while we expect 36 from a successful jet of the same total energy. However, it should be noted that we expect internal shocks to be important for successful jets. Considering our successful jet with $L_{\rm{iso}}=10^{52}\,\mathrm{erg\,s^{-1}}$, $T_j=100\,\mathrm{s}$, and $t_v=0.02\,\mathrm{s}$ (so that internal shocks occur at $r_s \approx r_\times$), we find that the total muon event number from the internal shock is $\sim$100 times larger than our estimates from the reverse shock~\cite{Meszaros:2001ms,Razzaque:2003uv,Razzaque:2004yv,Ando:2005xi}. This is due to less meson cooling in the cold jet, and we compare in Fig.~\ref{suppression2} the suppression factor for mesons in the cold jet and hot jet. It is clear that suppression in the cold jet is much weaker.

Despite being overwhelmed in terms of numbers, reverse shock neutrinos will arrive prior to neutrinos from internal shocks. M\'esz\'aros and Waxman~\cite{Meszaros:2001ms} studied precursor signals arising from internal shock accelerated protons interacting with photons at $r\sim10^{12}\,\mathrm{cm}$, while Razzaque, M\'esz\'aros, and Waxman~\cite{Razzaque:2003uv} extended this to include $r\sim r_{\rm He}$. Neutrinos from reverse shocks will precede even these precursor signals, provided the radius is large enough that the stellar material is thin to neutrinos. The temporal profile of neutrinos therefore offers the possibility to distinguish locations of particle acceleration for exceptionally close supernovae. The expected time difference between components is of order a few to tens seconds, which is within the time resolution of neutrino detectors. 

We classified jets by whether they break out of the star or not, depending on their durations. Choked jets are bright in neutrinos and dark in gamma rays, while successful jets are bright in neutrinos but may or may not be bright in gamma rays, depending on whether internal shocks are optically thick or not. A successful yet gamma dark jet may be typical for mildly relativistic jets, where the smaller $\Gamma_j$ implies internal shocks typically occur at smaller radii~\cite{Razzaque:2004yv,Ando:2005xi}. These could be distinguished from choked jets by the presence of an afterglow like radio signature, for example. 

In the current work we have assumed a type Ib progenitor. However, if the progenitor is a type Ic, the situation is slightly different. Having no He core, the core crossing time is shorter, $t_{\rm{CNO}} \sim 20$ s, for $L_{\rm iso} = 10^{52}$ erg s$^{-1}$, and choked jets will become rarer. This implies the relative importance of internal shocks over reverse shocks. For successful ultrarelativistic jets in type Ic progenitors, the neutrino precursor signal can be as strong as those discussed previously~\cite{Meszaros:2001ms,Razzaque:2003uv}. Our results will however apply well to type II progenitors.

The core-collapse supernova rate within 10 Mpc is $\sim$1--3 yr$^{-1}$~\cite{Ando:2005ka}, and therefore, one could test the possibility of particle acceleration in the reverse shocks in the next decade. In particular, correlating with the optical signal of the supernova explosions (not necessarily the jet-like signature)~\cite{Kowalski:2007xb} or MeV neutrino detections~\cite{Ando:2005ka} would greatly help enhance the significance of the signal. Examples of recent close supernovae include SN~2007gr in NGC~1058 (distance $\sim$10 Mpc)~\cite{Crockett:2007bv}, and a further list can be found in Ref.~\cite{Prieto:2007yb}. The actual rate could in fact be higher, as a systematic survey of all nearby galaxies has not been performed to date.

\section{Conclusions}

We showed that neutrino emission from reverse shock accelerated protons are strongly suppressed due to meson cooling in the hot jet and hot stellar material surrounding the reverse shock. Despite this suppression, we still expect a fairly promising number of events at IceCube, ranging from $2$ to $20$ for a close supernova of $10\,\mathrm{Mpc}$ hosting a choked mildly relativistic jet of total isotropic energy $10^{53}$--$10^{54}\,\mathrm{erg}$. We find that for a supernova of given energy $E_{\rm{iso}}$, neutrino emission is favored by high-Lorentz factor, low-luminosity, and long-duration jets, for which meson cooling is less severe.

Our result sets the scale for neutrino emission from choked jets. For choked jets to provide more neutrinos, internal shocks are required to occur in the unshocked region of the jet. However, this is geometrically constrained from above by the reverse shock, and from below by the fact that the stellar matter is opaque to high-energy neutrinos for small radii. 

Regarding neutrino precursor signals which precede GRB photons, we conclude that the reverse shock neutrinos likely plays a small role. However, they will precede those neutrinos from internal shocks, allowing the temporal profile of precursor neutrinos from a close supernova to provide an identification of proton acceleration at the reverse shock.


\acknowledgments

We are grateful to John Beacom for discussions and comments, and Peter M\'esz\'aros for reading the script. We also thank Ehud Nakar and Yudai Suwa for early discussions, and Hylke Koers for email discussions after initial submission. SA was supported by the Sherman Fairchild Foundation.

\appendix
\section{Particle Cooling Mechanisms}

We discuss cooling mechanisms in the shocked jet head, which act on the proton (Section IIIA, shown in Fig.~\ref{cooling}) and mesons (Section IIIC). We explicitly denote by primes ($\prime$) the jet head comoving frame. All numerical values are derived for the proton.

First, the same magnetic field responsible for electron synchrotron radiation also causes energy losses in protons and mesons, given by
\begin{eqnarray}
t^\prime_{sync}
&=&(6 \pi m_x^4 c^3)/(\sigma_T m_e^2 B^{\prime2} \epsilon_x^\prime) \nonumber \\&\approx& 0.04 \, \left(
\frac{ r_{\times,10.7}^2}
{\varepsilon_{B,-1}L_{\rm{iso},52}\epsilon_{p,0}^\prime} \right)
\left( \frac{\bar{\Gamma}_{h,1}}{\Gamma_{j,1}} \right)^2 \,\mathrm{s},
\end{eqnarray}
where $m_e$ is the electron mass, and $m_x$ and $\epsilon^\prime_x$ are the particle mass and energy. The numerical value is derived for the proton. Due to the strong dependency on the mass, $\propto m^4$, synchrotron cooling is most effective for the pion meson and least effective for the proton.

Second, protons and mesons lose energy by inverse Compton (IC) scattering. The most natural target photons are due to radiation cooling of ultrarelativistic electrons. Although a full treatment using a black-body spectrum is simple, a simplistic approximation using Eq.~(\ref{photons}) and Eq.~(\ref{photons2}) is sufficient to demonstrate that IC cooling is subdominant. Below a transition energy $m_p^2 c^4/T^\prime_\gamma \sim 7 \times 10^4\,\mathrm{GeV}$ (for the proton), inverse Compton scattering proceeds in the Thomson regime, and the cooling time scale is given
\begin{eqnarray}
t^{\prime}_{IC,Th}&=&(3m_x^4c^3)/(4\sigma_T m_e^2 \epsilon_x^\prime T^\prime_\gamma \bar{n}^\prime_\gamma) \nonumber \\
&\approx& 0.04 \, 
\left( \frac{ r_{\times,10.7}^2}
{\varepsilon_{e,-1}L_{\rm{iso},52}\epsilon_{p,0}^\prime} \right) 
\left( \frac{\bar{\Gamma}_{h,1}}{\Gamma_{j,1}} \right)^2 \,\mathrm{s}.
\end{eqnarray} 
Again, note the strong dependency on the particle mass. Above the transition energy, inverse compton scattering proceeds in the Klein-Nishina limit, where the cooling time scale becomes 
\begin{eqnarray}
t^{\prime}_{IC,KN}&=&(3 \epsilon_x^\prime T^\prime_\gamma)/(4 \sigma_T m_e^2 c^5 \bar{n}^\prime_\gamma) \nonumber \\
&\approx& 10^{-11}\, \left( 
\frac{r_{\times,10.7}\epsilon^\prime_{p,0}}
{\varepsilon_{e,-1}^{1/2}L_{\rm{iso},52}^{1/2}} \right) 
\left( \frac{\bar{\Gamma}_{h,1}}{\Gamma_{j,1}} \right)^{-1} \,\mathrm{s}. 
\end{eqnarray}

Third, we consider cooling by the Bethe-Heitler (BH) interaction ($p \gamma \to p e^+e^-$), which is important for protons. Our treatment is the same as that described in Ref.~\cite{Razzaque:2004yv}. The BH cross section rises logarithmically with energy, $\sigma_{BH}=\alpha r_e^2 \{(28/9)\mathrm{ln}[(2\epsilon^\prime_p \epsilon^\prime_\gamma)/(m_p m_e c^4)-106/9]\}$~\cite{Razzaque:2004yv}. The $e^\pm$ produced are at rest in the centre of mass frame of the proton and photon, and so the energy lost by the proton in each interaction is given by $\Delta \epsilon^\prime_p=2m_e c^2 \gamma^\prime_{c.m.}$, where $\gamma^\prime_{c.m.}=(\epsilon^\prime_p + \epsilon^\prime_\gamma)/(m_p^2 c^4 + 2 \epsilon^\prime_p \epsilon^\prime_\gamma)^{1/2}$ is the Lorentz factor of the centre of inertia as seen in the comoving frame. The energy loss rate is $d\epsilon^\prime_p/dt^\prime=n^\prime_\gamma c \sigma_{BH} \Delta \epsilon^\prime_p$, and thus the cooling time is $t^\prime_{BH}=\epsilon^\prime_p / (d\epsilon^\prime_p/dt^\prime) = \epsilon^\prime_p/(2n^\prime_\gamma c \sigma_{BH} m_e c^2 \gamma^\prime_{c.m.})$.

Lastly we discuss cooling due to collisions with protons. Assuming $20\%$ of the proton energy is lost in each collisions with a proton, the cooling time scale is
\begin{eqnarray}
t^\prime_{pp}&=&\epsilon^\prime_p/(c
\sigma_{pp}n^\prime_p \Delta \epsilon^\prime_p) \nonumber \\
&\approx& 1 \times 10^{-6}\, \left(
\frac{ r_{\times,10.7}^2 \Gamma_{j,1} }
  {L_{\rm{iso},52}} \right) \, \mathrm{s},
\end{eqnarray}
for the proton, where $\sigma_{pp}=5 \times 10^{-26}\,\mathrm{cm^2}$~\cite{Eidelman:2004wy}. Analogous equations hold for mesons, e.g.~Eq.~(\ref{mesonhadronic}) for the pion.

     
\end{document}